\documentclass[%
 reprint,onecolumn,
superscriptaddress, 
12pt, 
nofootinbib,
 amsmath,amssymb,
floatfix,
]{revtex4-2}

\usepackage{graphicx}
\usepackage{dcolumn}
\usepackage{bm}

\usepackage{cancel}
\usepackage{amsfonts}
\usepackage{xcolor}
\usepackage{tcolorbox}

\newcommand{\beqa}{\begin{eqnarray}}
\newcommand{\eeqa}{\end{eqnarray}}
\newcommand{\nn}{\nonumber}

\begin{document}

\title{Quenched Dipole Pairs in Viscous Fluid Membranes across the Saffman Crossover: Integrable Hamiltonian Dynamics }

\author{Satyagni Bhattacharya}
\affiliation{Birla Institute of Technology and Science, Pilani, Hyderabad Campus, Telangana 500078, India}

\author{Debdatta Dey}
\affiliation{Birla Institute of Technology and Science, Pilani, Hyderabad Campus, Telangana 500078, India}

\author{Samyak Jain}
\affiliation{University of Wisconsin-Madison, Department of Physics and the Wisconsin IceCube Particle Astrophysics Center}

\author{Yassir Khan}
\affiliation{Birla Institute of Technology and Science, Pilani, Hyderabad Campus, Telangana 500078, India}

\author{Tirthankar Mazumder}
\affiliation{Indian Institute of Technology Bombay, India}

\author{Aryaman Mihir Seth}
\affiliation{Stanford University, Stanford, CA 94305-6106}

\author{Nikhil Mogalapalli}
\affiliation{Birla Institute of Technology and Science, Pilani, Hyderabad Campus, Telangana 500078, India}

\author{Divyansh Tiwari}
\affiliation{Indian Institute of Technology Bombay, India}

\author{Pravallika Vemparala}
\affiliation{Massachusetts Institute of Technology, Cambridge, MA}

\author{Rickmoy Samanta}
\affiliation{Birla Institute of Technology and Science, Pilani, Hyderabad Campus, Telangana 500078, India}
\affiliation{Indian Statistical Institute, 203 B. T. Road, Kolkata 700108, India}
\affiliation{Indian Institute of Technology Kharagpur, West Bengal 721302, India}
\date{\today}
\begin{abstract}
We investigate an analytic theory of force-dipole hydrodynamics in a viscous membrane coupled to an infinite surrounding fluid, focusing on quenched (orientation-fixed) dipoles. While the single-dipole flow exhibits the known Saffman crossover from a near-field \(v\sim r^{-1}\) to a screened far-field \(v\sim r^{-2}\), we show that this crossover induces a qualitatively new reorganization of dipole-dipole interactions. For two identical quenched dipoles, the near-field dynamics is exactly solvable and effectively one-dimensional, with a fixed line of centers and linear evolution of the squared separation. In the far field, the system remains integrable but becomes intrinsically two-dimensional, with coupled radial and angular dynamics and an exact first integral. For pullers, the angular dynamics drives alignment toward an attracting manifold, leading to universal late-time collapse \(R\sim (t_c-t)^{1/3}\), in contrast to the near-field scaling \(R\sim (t_c-t)^{1/2}\). The Saffman crossover thus reorganizes the Hamiltonian phase-space structure of dipolar interactions and produces a transition from effectively one-dimensional to fully coupled dynamics, providing a minimal framework for aggregation in viscous fluid membranes.
\end{abstract}

\maketitle
\section{Introduction}

Hydrodynamics at low Reynolds number governs transport and interactions in a wide range of soft and biological systems, where viscous forces dominate and flow fields are long-ranged~\cite{purcell,lauga2009,lamb}. Fluid membranes provide a natural example for such dynamics, combining two-dimensional incompressible in-plane flow with momentum exchange with the surrounding three-dimensional solvent. This coupling introduces the Saffman-Delbr\"uck screening length, which separates logarithmic membrane flow at short distances from algebraically decaying flow at large distances~\cite{saff1,saff2}. The hydrodynamic response of flat membranes has been extensively characterized for passive inclusions~\cite{hughes,evans,lg96,staj,fischer,OppenheimerDiamant2009,OppenheimerDiamant2010,OppenheimerDiamant2011}.

In active systems, force-dipolar disturbances provide the leading far-field signature of force- and torque-free swimmers and inclusions~\cite{lauga2009}. In membranes, such dipoles offer a minimal description of active proteins and driven inclusions, giving rise to long-ranged interactions and collective dynamics~\cite{Graaf,mik,mnk,aggexp1,aggexp2}. Recent work has highlighted the role of curvature and odd viscosity in modifying these interactions~\cite{henlev2008,henlev2010,wg2012,wg2013,atzbergershape,atz2016,atz2018,atz2019,BagariaSamanta2022,JainSamanta2023,HosakaKomuraAndelman2021,Hosaka2023,hosaka2026,KrishnanOddViscosity2026}, while a complementary Hamiltonian perspective suggests that membrane-mediated dynamics may admit nontrivial reduced structure even in simple geometries~\cite{KrishnanSamanta2026}.

Here we revisit the simplest setting of force-dipole hydrodynamics in a flat membrane coupled to an infinite subphase and analyze the interaction of dipole pairs across the Saffman crossover. We focus on the quenched limit, in which dipole orientations are fixed. This regime is relevant when dipole axes are externally constrained or when orientational relaxation is slow compared to translational motion, as in anchored ``shakers'' and strongly aligned active suspensions~\cite{Shoham2023}. It is also analytically advantageous, as fixed orientations yield a consistent position-based Hamiltonian description, as we show in this work.

We derive explicit expressions for the dipolar velocity and vorticity fields and show that hydrodynamic screening induces a qualitative change in flow structure, from a purely radial near-field flow to a screened far-field flow with nonradial components. We then analyze the interaction of two identical quenched dipoles. In the near zone, the dynamics is exactly solvable and effectively one-dimensional, with a fixed line of dipole-centers and linear evolution of the squared separation. In the far zone, the system remains integrable but becomes intrinsically two-dimensional, with coupled radial and angular dynamics and an exact first integral. For pullers, the angular dynamics drives the system toward an aligned attracting manifold, leading to universal late-time collapse with \(R\sim (t_c-t)^{1/3}\), in contrast to the near-field scaling \(R\sim (t_c-t)^{1/2}\). These results show that the Saffman crossover reorganizes not only the spatial decay of membrane flows but also the Hamiltonian phase-space structure of dipolar interactions, providing a minimal analytic framework for aggregation and transport in active membrane systems.
\paragraph*{Organization of the paper.}
We begin in Sec.~\ref{mflow} by formulating the hydrodynamics of a free viscous membrane coupled to an infinite surrounding fluid and deriving the corresponding membrane Green's function and Stokeslet response. In Sec.~\ref{dipoleflows}, we construct the flow generated by an in-plane force dipole and obtain explicit near- and far-field asymptotic forms for the velocity and vorticity across the Saffman crossover. We then turn in Sec.~\ref{ans} to the interaction of two identical quenched, coaligned dipoles. In the near zone, we show that the relative dynamics is exactly solvable and effectively one-dimensional, with a fixed line of centers and linear evolution of the squared separation, while in the far zone we derive an integrable but intrinsically two-dimensional reduced dynamics with coupled radial and angular motion, an exact first integral, and universal late-time cubic collapse. In Sec.~\ref{sumarystrm}, we summarize the asymptotic flow and streamfunction structure in a compact form. We then recast the two-body problem in Hamiltonian form in Sec.~\ref{sec:two_body_ham}, showing that the Saffman crossover reorganizes the phase-space structure from a degenerate near-field dynamics with frozen angle to a fully coupled far-field dynamics with an attracting manifold. Finally, in Sec.~\ref{cncl} we summarize the main results and discuss possible extensions to orientable dipoles, many-body interactions, and more complex membrane environments.
\section {Membrane flows}
\label{mflow}
We revisit some results from analysis of force dipoles in flat membranes with infinite subphase \cite{evans,lg96,staj,fischer, mnk}.  Let us first consider  a free flat membrane situated at $z=0$ and infinitely extended along $x,y$. We denote the membrane pressure by $p$, the  viscosity of the 2D membrane fluid (incompressible) $\eta_{2D}$, $v$ is the in-plane 2D velocity along the membrane,  $\eta_\pm$ denotes the 3D viscosities of external solvents, $v_\pm$ represent the velocities of the 3D external fluids above(+) and below(-) the membrane, respectively. The coupled set of hydrodynamic equations at low Reynolds is given by
\beqa
&\nabla_{s}.\vec{v}=0 \nn\\
& \eta_{s} \nabla_{2D}^2 v_\alpha -\partial_\alpha p_{2D} + \left( \sigma_{\alpha z}^+ - \sigma_{\alpha z}^-\right)|_{z\rightarrow 0} +f_\alpha \delta^2(\bm{r})=0\nn\\
& \nabla.\vec{v}_\pm =0, ~~~ \eta_{\pm} \nabla^2 \vec{v}_\pm = -\nabla p_\pm
\label{plane_hydro}
\eeqa
where the index $\alpha$ denotes the in-plane membrane co-ordinates, $\nabla_{2D}$ is the two dimensional derivative and the bulk fluid stress tensor is given by
\beqa
\sigma_{ij}^\pm = \eta_\pm \left( \partial_i v^\pm_j + \partial_j v^\pm_i\right)
\eeqa
along with the usual no-slip boundary conditions at the membrane surface
\beqa
\vec{v} _\pm |_{z=0}= \vec{v}
\eeqa
Performing a 2D Fourier transform on the incompressibility and momentum balance equation in Eq.(\ref{plane_hydro}), and assuming $\eta_\pm =\eta$, we arrive at
\beqa
&q^\alpha v_\alpha =0, \nn\\
&-\eta_{s} q^2 v_\alpha(q) - i  q_\alpha p - 2 \eta~ q~ v_\alpha(q) +f_\alpha =0.
\label{mbl}
\eeqa
Contracting the second equation with $q^\alpha$ and using the incompressibility relation,  we get
\beqa
p= \frac{f_\alpha q^\alpha}{i q^2}
\eeqa 
Plugging the membrane pressure back into the momentum balance equation in Eq.(\ref{mbl}), we get 
\beqa
v^{Stokeslet}_i(q) \equiv G^{free}_{i j}(q)~f_j= \frac{\delta_{ij} - \frac{q_i q_j}{q^2} }{\eta_{s}~ q (q+ \lambda^{-1})}f_j
\label{gfree}
\eeqa
where the length scale $\lambda =\frac{\eta_{s}}{\eta_+ +\eta_-}$ is the Saffman length and $G^{free}_{ij}(q)$ is the Greens function for the free membrane in Fourier space. Performing the inverse Fourier transform (please see Appendix \ref{app1} for a self-contained derivation), the expression for the velocity field in real space is given by
\beqa
v^{\text{Stokeslet}}_i
:=
G^{free}_{ij} f_j
=
\left[
A_0(r)\,\delta_{ij}
+
A_1(r)\,\hat r_i \hat r_j
\right] f_j,
\qquad
\hat r_i = \frac{r_i}{r},
\label{plane_pf_main}
\eeqa
where the scalar kernels are given by
\beqa
A_0(r)
&=&
\frac{1}{4\pi\eta_{s}}
\left[
\pi H_0\!\left(\frac{r}{\lambda}\right)
-\pi \frac{H_1\!\left(\frac{r}{\lambda}\right)}{r/\lambda}
+\frac{2}{(r/\lambda)^2}
-\pi \frac{Y_0\!\left(\frac{r}{\lambda}\right)-Y_2\!\left(\frac{r}{\lambda}\right)}{2}
\right],
\nn\\
A_1(r)
&=&
\frac{1}{4\pi\eta_{s}}
\left[
-\pi H_0\!\left(\frac{r}{\lambda}\right)
+2\pi \frac{H_1\!\left(\frac{r}{\lambda}\right)}{r/\lambda}
-\frac{4}{(r/\lambda)^2}
-\pi Y_2\!\left(\frac{r}{\lambda}\right)
\right].
\eeqa
Here $H_n$ denotes the Struve function of order $n$ and $Y_n$ the Bessel function of the second kind of order $n$.

It is useful to consider the asymptotic limits:
\beqa
A_0(r)
~\rightarrow~
\frac{1}{4\pi\eta_{s}}
\begin{cases}
\log\!\left(\frac{\lambda}{r}\right), & r\ll \lambda,\\[6pt]
2\frac{\lambda^2}{r^2}, & r\gg \lambda,
\end{cases}
\eeqa
and
\beqa
A_1(r)
~\rightarrow~
\frac{1}{4\pi\eta_{s}}
\begin{cases}
1, & r\ll \lambda,\\[6pt]
\displaystyle \frac{2\lambda}{r} - \frac{4\lambda^2}{r^2}, & r\gg \lambda.
\end{cases}
\eeqa
Thus, the Green's function exhibits the familiar logarithmic behavior of 2D Stokes flow at short distances $r\ll\lambda$, which is regulated at large distances by momentum leakage into the surrounding fluid, leading to algebraically decaying flows for $r\gg\lambda$. The velocity field generated by a force dipole follows from the directional derivative of the Stokeslet along the dipole orientation. For the free membrane one obtains
\beqa
[v_i]^{\text{dipole}}
=
\sigma\, \hat d_k\, \partial_{x_k}
\left[
G^{free}_{ij}\, \hat d_j
\right],
\label{vdp_free}
\eeqa
where $G^{free}_{ij}$ is given by Eq.~(\ref{plane_pf_main}) and 
$\sigma = |\mathbf f|\, dL$ is held fixed in the point-dipole limit 
$dL\to 0$, $|\mathbf f|\to \infty$. Here $\partial_{x_k}$ denotes differentiation 
with respect to the observation coordinate.

In the present convention, the sign of $\sigma$ determines the dipolar character: 
$\sigma>0$ corresponds to a puller, with fluid drawn inward along the dipole axis 
and expelled transversely, while $\sigma<0$ corresponds to a pusher, with fluid 
expelled along the dipole axis and drawn inward from the sides.

\section{Dipolar flows in a free membrane: near- and far-field structure}
\label{dipoleflows}
\begin{figure}[t]
\centering
\includegraphics[width=\columnwidth]{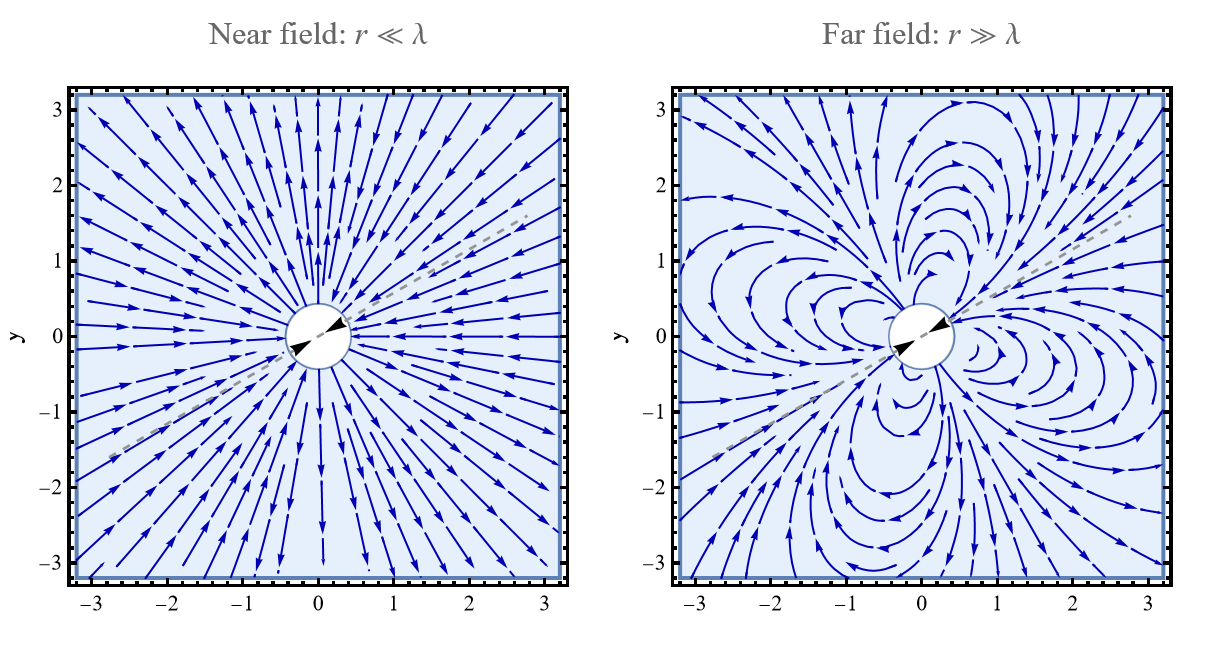}
\caption{
Near- and far-field velocity fields generated by a single force dipole (puller) in a viscous membrane. 
The dipole axis (dashed line) is oriented at an angle $\alpha=\pi/6$ with respect to the $x$-axis, with inward arrows indicating its contractile nature. 
\textbf{Left:} Near-field regime ($r \ll \lambda$), where the flow is purely radial and anisotropic, with no rotational component. 
\textbf{Right:} Far-field regime ($r \gg \lambda$), where hydrodynamic screening at the Saffman length $\lambda$ produces non-radial components and a pronounced rotational structure in the streamlines. 
The central disk denotes the excluded core region.
}
\label{fig:dipole_flow}
\end{figure}
The velocity field generated by a force dipole in a membrane can be obtained by differentiating the Stokeslet along the dipole axis. For a dipole of strength $\sigma$ and orientation $\hat{\mathbf d}$, the flow takes the form
\begin{equation}
v_i(\mathbf r)
=
\sigma\, \hat d_k\, \partial_k G_{ij}(\mathbf r)\, \hat d_j,
\end{equation}
where $G_{ij}(\mathbf r)$ is the hydrodynamic Green’s tensor. In an isotropic membrane, the Green’s function admits the decomposition
\begin{equation}
G_{ij}(\mathbf r)
=
A_0(r)\,\delta_{ij}
+
A_1(r)\,\hat r_i \hat r_j
+
A_2(r)\,\varepsilon_{ij},
\qquad
r = |\mathbf r|,\quad \hat r_i = r_i/r.
\end{equation}
For a free membrane with an infinite subphase, the antisymmetric contribution vanishes, $A_2(r)=0$, and the dipolar velocity field reduces to
\begin{equation}
\mathbf v(\mathbf r)
=
\sigma
\left[
\gamma\!\left(A_0'(r)+\frac{A_1(r)}{r}\right)\hat{\mathbf d}
+
\left(
\frac{A_1(r)}{r}
+
\gamma^2\!\left(A_1'(r)-\frac{2A_1(r)}{r}\right)
\right)\hat{\mathbf r}
\right],
\end{equation}
while the associated vorticity field is
\begin{equation}
\omega(\mathbf r)
=
\sigma\,\gamma\beta
\left(
A_0''(r)
-\frac{A_0'(r)}{r}
-\frac{A_1'(r)}{r}
+\frac{2A_1(r)}{r^2}
\right).
\end{equation}
Here the angular dependence is encoded through the projections
\begin{equation}
\gamma = \hat{\mathbf d}\cdot \hat{\mathbf r}, 
\qquad
\beta = \varepsilon_{ij} d_j \hat r_i,
\qquad
\gamma^2+\beta^2=1.
\end{equation}

The structure of the flow is controlled by the asymptotic behavior of the scalar kernels $A_0(r)$ and $A_1(r)$, which exhibit a crossover at the Saffman length $\lambda = \eta_s/(2\eta_f)$. In the near-field regime $r\ll \lambda$, the Green’s function reduces to the two-dimensional Stokes form,
\begin{equation}
A_0(r)\sim \frac{1}{4\pi\eta_s}\log\!\left(\frac{\lambda}{r}\right),
\qquad
A_1(r)\sim \frac{1}{4\pi\eta_s}.
\end{equation}
Substituting into the general expressions yields a cancellation of the longitudinal component, giving
\begin{equation}
\mathbf v_{\rm near}(\mathbf r)
=
\frac{\sigma}{4\pi\eta_s}
\frac{1-2\gamma^2}{r}\,\hat{\mathbf r},
\qquad
\omega_{\rm near}(\mathbf r)
=
\frac{\sigma}{\pi\eta_s}
\frac{\gamma\beta}{r^2}.
\end{equation}
Thus, the near-field dipolar flow is purely radial at leading order, with a quadrupolar angular structure and a logarithmic origin inherited from two-dimensional hydrodynamics.

In the far-field regime $r\gg \lambda$, the Green’s function is screened by momentum leakage into the surrounding fluid, with asymptotic forms
\begin{equation}
A_0(r)\sim \frac{1}{4\pi\eta_s}\,2\frac{\lambda^2}{r^2},
\qquad
A_1(r)\sim \frac{1}{4\pi\eta_s}\left(\frac{2\lambda}{r}-\frac{4\lambda^2}{r^2}\right).
\end{equation}
Substitution into the dipole expressions gives
\begin{equation}
\mathbf v_{\rm far}(\mathbf r)
=
\sigma
\left[
\frac{\gamma \lambda (r-4\lambda)}{2\pi \eta_s r^3}\,\hat{\mathbf d}
+
\frac{\lambda\bigl(r-3r\gamma^2+(-2+8\gamma^2)\lambda\bigr)}{2\pi \eta_s r^3}\,\hat{\mathbf r}
\right],
\end{equation}
\begin{equation}
\omega_{\rm far}(\mathbf r)
=
\frac{3\sigma\gamma\beta\lambda}{2\pi\eta_s r^3}.
\end{equation}
Retaining only the leading far-field contribution, one obtains the screened dipolar flow
\begin{equation}
\mathbf v_{\rm far}(\mathbf r)
=
\frac{\sigma\lambda}{2\pi\eta_s r^2}
\left[
\gamma\,\hat{\mathbf d}
+
(1-3\gamma^2)\,\hat{\mathbf r}
\right]
+
\mathcal{O}\!\left(\frac{\lambda^2}{r^3}\right).
\end{equation}
These results demonstrate a crossover from a logarithmic two-dimensional dipolar flow at short distances to a screened, algebraically decaying flow at large distances, with the angular structure entirely governed by the projections $\gamma$ and $\beta$.
\section{ Exact Analytic Solutions of Quenched and co-aligned dipole pair dynamics}
\label{ans}
\begin{figure}[t]
\centering
\includegraphics[width=0.65\linewidth]{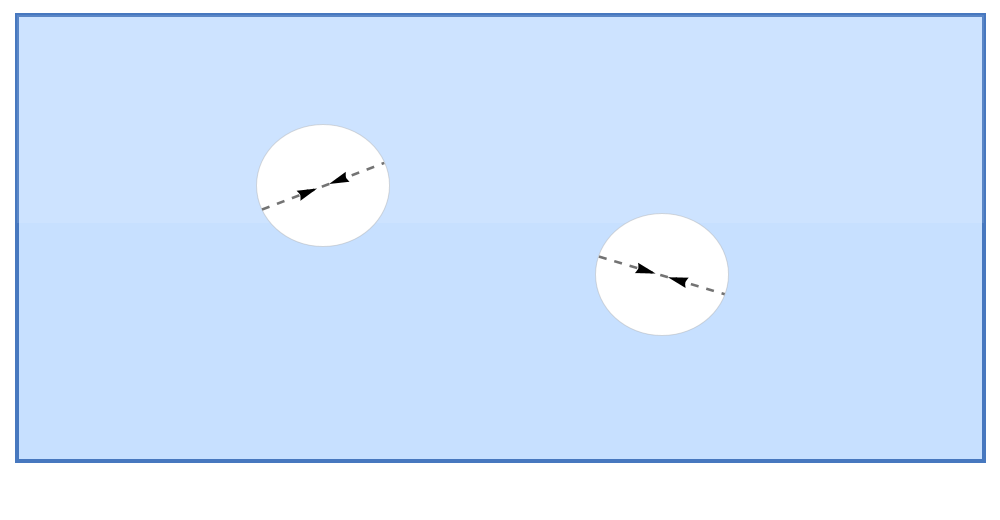}
\caption{
Schematic of two identical force dipoles (pullers) embedded in a viscous fluid membrane. Each dipole is characterized by a fixed orientation $\hat{\mathbf d}$ (quenched limit), indicated by the dashed axis and arrows. The surrounding medium provides hydrodynamic coupling between the dipoles, leading to collective motion governed by membrane-mediated interactions.
}
\label{fig:two_puller_schematic}
\end{figure}
\begin{figure}[t]
\centering
\includegraphics[width=0.7\linewidth]{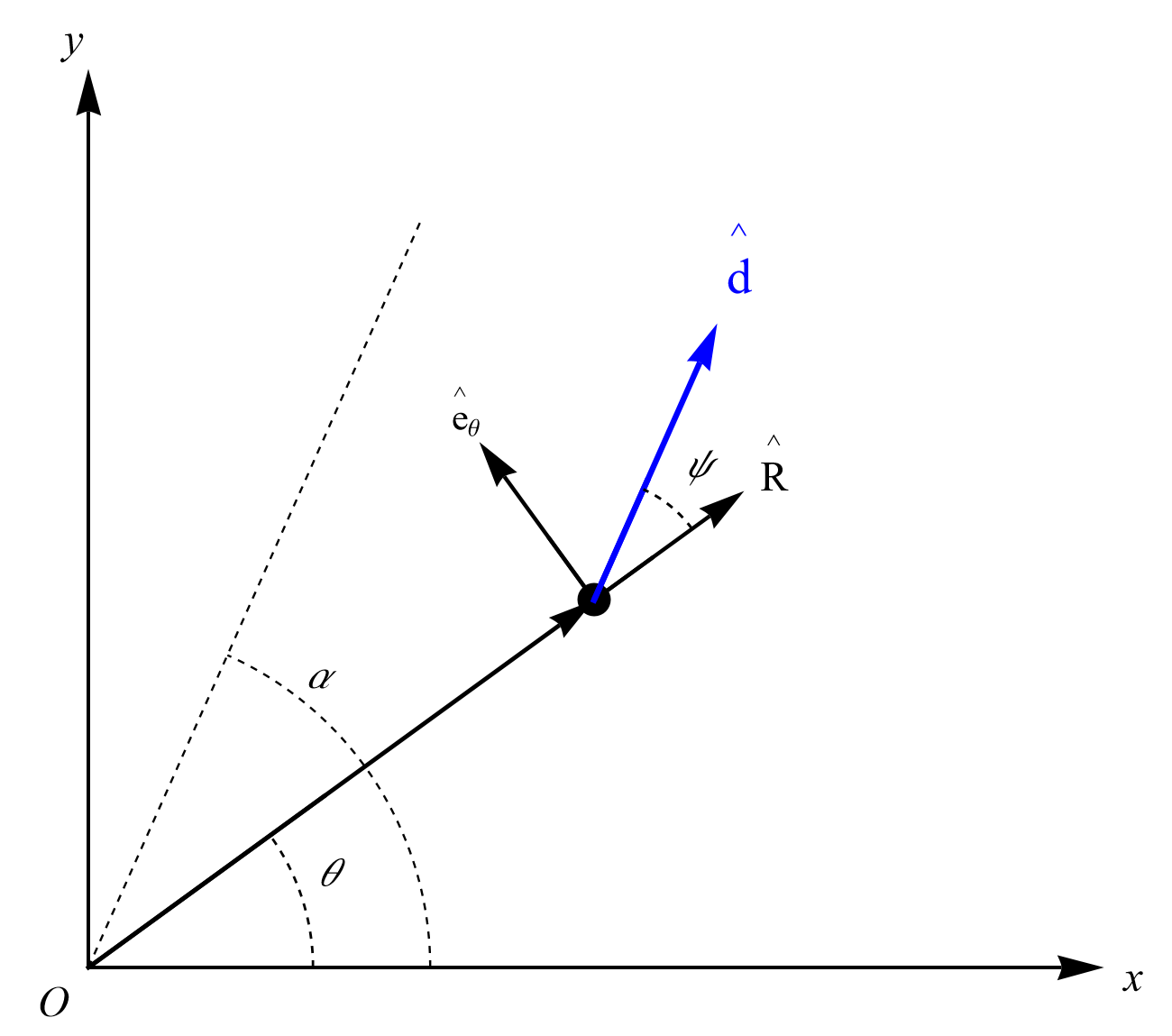}
\caption{Geometry of two aligned quenched dipoles in the plane. 
The relative position vector $\mathbf R = \mathbf r_2 - \mathbf r_1$ defines the unit vector $\hat{\mathbf R}$ and its polar angle $\theta$ with respect to the laboratory $x$-axis. The dipole orientation $\hat{\mathbf d}$ is fixed (quenched) and makes a lab-frame angle $\alpha$ with the $x$-axis.  The relative angle between the dipole axis and the line of centers is denoted by $\psi$, defined through $\hat{\mathbf d}\cdot \hat{\mathbf R} = \cos\psi$. The local polar basis at the dipole location is given by $(\hat{\mathbf R}, \hat{\boldsymbol\theta})$, where $\hat{\boldsymbol\theta}$ is perpendicular to $\hat{\mathbf R}$. In this basis, the dipole orientation decomposes as
$\hat{\mathbf d} = \cos\psi\,\hat{\mathbf R} + \sin\psi\,\hat{\boldsymbol\theta}$.
The three angles are related by $\alpha = \theta + \psi$. Since $\alpha$ is fixed in the quenched limit, the rotation of the line of centers implies $\dot{\theta} = -\dot{\psi}$.
}
\label{fig:angles}
\end{figure}
\begin{figure}[t]
\centering
\includegraphics[width=0.7\linewidth]{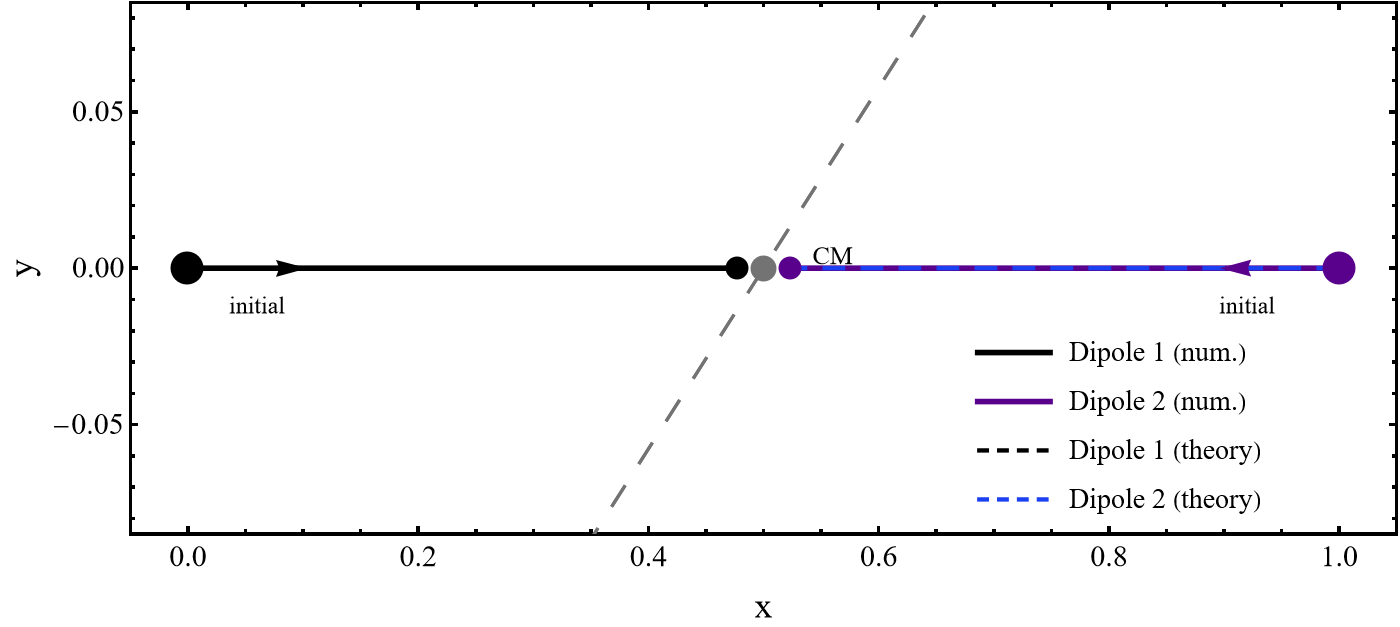}
\caption{
Near-zone trajectories of two aligned quenched dipoles (pullers) with initial positions $\mathbf r_1(0)=(0,0)$ and $\mathbf r_2(0)=(1,0)$, and dipole orientation $\alpha_d=\pi/6$. Solid lines denote numerical solutions, while dashed lines show the exact analytic solution. The motion remains strictly collinear along the initial line of centers, with the center of mass fixed and the dipoles undergoing monotonic approach leading to finite-time collapse. The dashed gray line marks the fixed dipole axis $\hat{\mathbf d}$, which remains constant in the quenched limit and does not influence the direction of motion in the near zone.
}
\label{fig:quenchnearzone}
\end{figure}
\begin{figure}[t]
\centering
\includegraphics[width=0.7\linewidth]{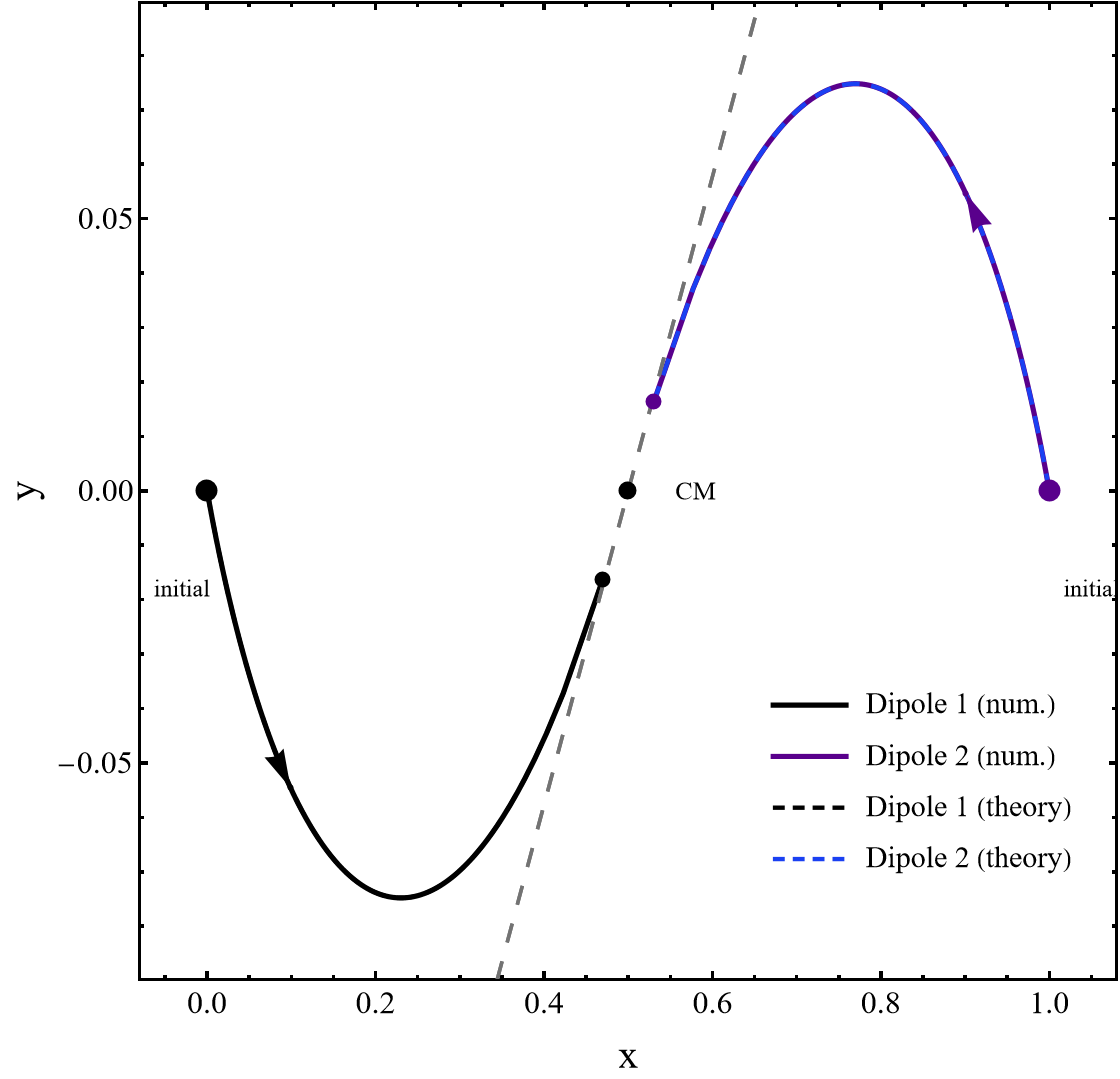}
\caption{
Far-zone trajectories of two aligned quenched dipoles (pullers) with initial conditions 
$\mathbf r_1(0)=(0,0)$, $\mathbf r_2(0)=(1,0)$ and dipole orientation 
$\alpha_d=\pi/6$. Solid lines denote numerical solutions of the reduced dynamical system, while dashed lines show the analytic parametric solution derived from the implicit evolution of $\psi(t)$. The center of mass remains fixed, and the two dipoles undergo a combined rotational and radial motion. The  agreement demonstrates the validity of the far-zone reduction up to times close to the finite-time collision. The dashed gray line indicates the fixed dipole axis $\hat{\mathbf d}$, which remains constant in the quenched limit.
}
\label{fig:quenchfarzone}
\end{figure}
\subsection{Near-zone dynamics of two aligned quenched dipoles}

We consider two identical force dipoles in a free membrane in the near zone, \(R\ll \lambda\), with fixed and aligned orientations \(\hat{\mathbf d}_1=\hat{\mathbf d}_2=\hat{\mathbf d}\). Writing the relative coordinate as
\begin{equation}
\mathbf R=\mathbf r_2-\mathbf r_1,\qquad
R=|\mathbf R|,\qquad
\hat{\mathbf R}=\mathbf R/R,
\end{equation}
the near-field flow generated by a single dipole is
\begin{equation}
\mathbf v(\mathbf r)=\frac{\sigma}{4\pi\eta_s}\frac{1-2\gamma^2}{r}\,\hat{\mathbf r},
\qquad
\gamma=\hat{\mathbf d}\cdot\hat{\mathbf r}.
\end{equation}
Evaluating this at the position of the other dipole gives
\begin{equation}
\dot{\mathbf R}
=
\dot{\mathbf r}_2-\dot{\mathbf r}_1
=
\frac{\sigma}{2\pi\eta_s}\frac{1-2\gamma^2}{R}\,\hat{\mathbf R},
\qquad
\gamma=\hat{\mathbf d}\cdot\hat{\mathbf R}.
\label{eq:near_rel_vec}
\end{equation}
The motion is therefore purely radial: \(\dot{\mathbf R}\parallel \hat{\mathbf R}\), so \(\dot{\hat{\mathbf R}}=0\), and hence \(\gamma\) is constant. Projecting Eq.~\eqref{eq:near_rel_vec} along \(\hat{\mathbf R}\) yields
\begin{equation}
\dot R=\frac{\sigma}{2\pi\eta_s}\frac{1-2\gamma^2}{R},
\end{equation}
which integrates exactly to
\begin{equation}
R^2(t)=R_0^2+\frac{\sigma}{\pi\eta_s}(1-2\gamma^2)(t-t_0).
\label{eq:near_R2_law}
\end{equation}
Thus the relative trajectory remains on the initial line of centers,
\begin{equation}
\mathbf R(t)=
\sqrt{R_0^2+\frac{\sigma}{\pi\eta_s}(1-2\gamma^2)(t-t_0)}\,\hat{\mathbf R}_0.
\end{equation}

The sign of \(1-2\gamma^2\) determines whether the interaction is attractive or repulsive. Writing \(\gamma=\cos\psi\), where \(\psi\) is the angle between \(\hat{\mathbf d}\) and \(\hat{\mathbf R}\), one has \(1-2\gamma^2=-\cos 2\psi\). Thus, for pullers (\(\sigma>0\)), configurations with \(\psi<\pi/4\) (\(\gamma^2>1/2\)) are attractive, whereas those with \(\psi>\pi/4\) (\(\gamma^2<1/2\)) are repulsive; the marginal case \(\psi=\pi/4\) gives vanishing leading-order interaction. In particular, when \(\hat{\mathbf d}\parallel \hat{\mathbf R}\) (\(\psi=0\), \(\gamma=1\)), Eq.~\eqref{eq:near_R2_law} gives finite-time collapse,
\begin{equation}
R^2(t)=R_0^2-\frac{\sigma}{\pi\eta_s}(t-t_0),
\qquad
t_c=t_0+\frac{\pi\eta_s R_0^2}{\sigma},
\end{equation}
whereas for \(\hat{\mathbf d}\perp \hat{\mathbf R}\) (\(\psi=\pi/2\), \(\gamma=0\)) the separation grows monotonically. The center of mass remains fixed, so the individual trajectories follow as
\begin{equation}
\mathbf r_1(t)=\mathbf r_{\rm cm}-\frac{1}{2}\mathbf R(t),
\qquad
\mathbf r_2(t)=\mathbf r_{\rm cm}+\frac{1}{2}\mathbf R(t).
\end{equation}
The near-zone problem is therefore exactly integrable and effectively one-dimensional: the pair moves on a fixed straight line, while the squared separation evolves linearly in time. Figure~\ref{fig:quenchnearzone} confirms this prediction through agreement between the analytic solution and direct numerical integration.

\subsection{Far-zone dynamics of two aligned quenched dipoles}

We next consider two identical force dipoles in the far zone, \(R\gg \lambda\), with fixed and aligned orientations \(\hat{\mathbf d}_1=\hat{\mathbf d}_2=\hat{\mathbf d}\). 
The leading far-field flow generated by a single dipole is
\begin{equation}
\mathbf v(\mathbf r)=
\frac{\sigma\lambda}{2\pi\eta_s r^2}
\left[
\gamma\,\hat{\mathbf d}
+
(1-3\gamma^2)\hat{\mathbf r}
\right],
\qquad
\gamma=\hat{\mathbf d}\cdot\hat{\mathbf r}.
\end{equation}
Evaluating this at the position of the other dipole gives the relative equation
\begin{equation}
\dot{\mathbf R}
=
\frac{\sigma\lambda}{2\pi\eta_s R^2}
\left[
2\gamma\,\hat{\mathbf d}
+
(2-6\gamma^2)\hat{\mathbf R}
\right],
\qquad
\gamma=\hat{\mathbf d}\cdot\hat{\mathbf R}.
\label{eq:far_rel_vec}
\end{equation}
Unlike the near-zone case, Eq.~\eqref{eq:far_rel_vec} contains a component along the fixed dipole axis and is therefore not purely radial. Introducing the relative angle \(\psi\) through
\begin{equation}
\gamma=\cos\psi=\hat{\mathbf d}\cdot\hat{\mathbf R},
\end{equation}
and writing
\begin{equation}
\dot{\mathbf R}=\dot R\,\hat{\mathbf R}-R\dot\psi\,\hat{\boldsymbol\theta},
\end{equation}
one obtains the coupled system
\begin{equation}
\dot R=
-\frac{\sigma\lambda}{\pi\eta_s}\frac{\cos 2\psi}{R^2},
\qquad
\dot\psi=
-\frac{\sigma\lambda}{2\pi\eta_s}\frac{\sin 2\psi}{R^3}.
\label{eq:far_Rpsi}
\end{equation}
Thus the far-zone dynamics remains anisotropic, but now the line of centers rotates in time.

Equation~\eqref{eq:far_Rpsi} admits a first integral. Dividing the two equations gives
\begin{equation}
\frac{dR}{d\psi}=2R\cot 2\psi,
\end{equation}
which integrates to
\begin{equation}
R=C\sin 2\psi
=
R_0\frac{\sin 2\psi}{\sin 2\psi_0},
\label{eq:far_orbit}
\end{equation}
where \(R_0=R(t_0)\) and \(\psi_0=\psi(t_0)\). Substituting Eq.~\eqref{eq:far_orbit} into Eq.~\eqref{eq:far_Rpsi} yields the implicit evolution law
\begin{equation}
(\psi-\psi_0)-\frac{1}{4}\bigl(\sin 4\psi-\sin 4\psi_0\bigr)
=
-\frac{\sigma\lambda\,\sin^3(2\psi_0)}{\pi\eta_s R_0^3}(t-t_0).
\label{eq:far_implicit}
\end{equation}
Equations~\eqref{eq:far_orbit} and \eqref{eq:far_implicit} therefore provide the complete far-zone solution in implicit form. The problem remains integrable, but in contrast to the near zone the motion is intrinsically two-dimensional.

The sign of \(\cos 2\psi\) determines whether the interaction is attractive or repulsive. For pullers (\(\sigma>0\)), configurations with \(\psi<\pi/4\) are attractive, those with \(\psi>\pi/4\) are repulsive, and \(\psi=\pi/4\) gives vanishing radial motion at leading order. Two invariant limits are especially simple. For \(\psi=0\), corresponding to \(\hat{\mathbf d}\parallel \hat{\mathbf R}\), the angular dynamics vanishes and Eq.~\eqref{eq:far_Rpsi} reduces to
\begin{equation}
\dot R=-\frac{\sigma\lambda}{\pi\eta_s R^2},
\qquad
R^3(t)=R_0^3-\frac{3\sigma\lambda}{\pi\eta_s}(t-t_0),
\end{equation}
which gives finite-time collapse for pullers. For \(\psi=\pi/2\), corresponding to \(\hat{\mathbf d}\perp \hat{\mathbf R}\), one similarly finds
\begin{equation}
\dot R=+\frac{\sigma\lambda}{\pi\eta_s R^2},
\qquad
R^3(t)=R_0^3+\frac{3\sigma\lambda}{\pi\eta_s}(t-t_0),
\end{equation}
corresponding to monotonic separation.

For the aligned quenched pair considered here, the center of mass remains fixed, so the individual trajectories follow from
\begin{equation}
\mathbf r_1(t)=\mathbf r_{\rm cm}-\frac{1}{2}\mathbf R(t),
\qquad
\mathbf r_2(t)=\mathbf r_{\rm cm}+\frac{1}{2}\mathbf R(t),
\end{equation}
with
\begin{equation}
\mathbf R(t)=
R_0\frac{\sin 2\psi(t)}{\sin 2\psi_0}
\bigl(\cos(\alpha_d-\psi(t)),\,\sin(\alpha_d-\psi(t))\bigr),
\end{equation}
where \(\alpha_d\) is the fixed dipole angle in the laboratory frame and \(\psi(t)\) is determined by Eq.~\eqref{eq:far_implicit}. The far-zone dynamics thus exhibits a clear qualitative crossover from the straight-line, purely radial near-zone motion to curved trajectories with coupled radial and angular evolution.
A further simplification emerges in the late-time regime approaching collapse. Expanding the implicit solution~\eqref{eq:far_implicit} for small $\psi$ gives
\begin{equation}
\psi - \frac{1}{4}\sin(4\psi)
=
\frac{8}{3}\psi^3 + O(\psi^5),
\end{equation}
which implies
\begin{equation}
t_c - t
=
\frac{8\pi\eta_s R_0^3}{3\sigma\lambda\,\sin^3(2\psi_0)}\,\psi^3
+ O(\psi^5),
\end{equation}
where $t_c$ denotes the finite collapse time. Thus $\psi \sim (t_c - t)^{1/3}$ as $t \to t_c$. Using the radial relation
\begin{equation}
R = R_0 \frac{\sin(2\psi)}{\sin(2\psi_0)}
\simeq \frac{2R_0}{\sin(2\psi_0)}\,\psi
\qquad (\psi \to 0),
\end{equation}
we obtain the  scaling law
\begin{equation}
R^3(t)
\sim
\frac{3\sigma\lambda}{\pi\eta_s}\,(t_c - t),
\end{equation}
independent of initial conditions. Although the far-field dynamics is intrinsically two-dimensional, the angular evolution drives the system toward the aligned configuration $\psi \to 0$. The late-time behavior is therefore governed by a cubic collapse law, $R \sim (t_c - t)^{1/3}$, with the aligned solution acting as an attracting manifold that controls the asymptotic aggregation dynamics.\\
To understand the angular dynamics, we linearize the leading-order far-field equations about $\psi=0$. Using $\sin(2\psi)\simeq 2\psi$, we obtain
\begin{equation}
\dot\psi
\simeq
-\frac{\lambda\sigma}{\pi\eta_s}\frac{\psi}{r_{12}^3},
\end{equation}
so that, for pullers ($\sigma>0$), $\psi=0$ is a linearly stable fixed point. In contrast, writing $\psi=\pi/2+\varepsilon$ and using $\sin(2\psi)\simeq -2\varepsilon$, we find
\begin{equation}
\dot\psi
\simeq
+\frac{\lambda\sigma}{\pi\eta_s}\frac{\varepsilon}{r_{12}^3},
\end{equation}
showing that the perpendicular configuration is unstable. 

These results imply that, in the far-field regime, the angular dynamics generically drives the system toward alignment with the dipole axis, $\psi\to 0$. Once the trajectory enters the sector $|\psi|<\pi/4$, one has $\cos(2\psi)>0$, and hence $\dot r_{12}<0$, so the separation decreases monotonically. Except for initial conditions lying exactly on the radial null directions $\psi=\pi/4+n\pi/2$ or the unstable perpendicular configuration, the system therefore evolves into an aggregating regime.\\
For pushers, writing \(\sigma=-|\sigma|\), the leading-order far-field system becomes
\[
\dot R=\frac{\lambda |\sigma|}{\pi\eta_s}\frac{\cos 2\psi}{R^2},
\qquad
\dot\psi=\frac{\lambda |\sigma|}{2\pi\eta_s}\frac{\sin 2\psi}{R^3}.
\]
Linearization shows that \(\psi=0\) is unstable, while \(\psi=\pi/2\) is stable. Defining \(\varepsilon=\pi/2-\psi\), one obtains
\[
\dot R=
-\frac{\lambda |\sigma|}{\pi\eta_s}\frac{\cos 2\varepsilon}{R^2},
\qquad
\dot\varepsilon=
-\frac{\lambda |\sigma|}{2\pi\eta_s}\frac{\sin 2\varepsilon}{R^3},
\]
which is identical in form to the puller problem. Thus pushers are driven toward the perpendicular manifold \(\psi\to\pi/2\), and along this stable branch
\[
R^3(t)=R_0^3-\frac{3\lambda |\sigma|}{\pi\eta_s}(t-t_0),
\]
so the same  cubic collapse law holds, with collapse controlled by the perpendicular rather than the aligned configuration.
\section{Summary of flows and stream functions in a viscous membrane: near-far crossover}
\label{sumarystrm}
We consider the velocity field generated by an in-plane force dipole of strength $\sigma$ and orientation angle $\alpha$ in a two-dimensional viscous membrane of viscosity $\eta_s$. In polar coordinates $(r,\theta)$ centered at the dipole, the flow exhibits distinct asymptotic forms separated by the Saffman-Delbr\"uck length $\lambda$.

In the near field ($r\ll\lambda$), the flow is purely radial,
\begin{equation}
v_r^{\rm near}
=
-\frac{\sigma}{4\pi\eta_s}
\frac{\cos\!\big[2(\alpha-\theta)\big]}{r},
\qquad
v_\theta^{\rm near}=0,
\end{equation}
yet it possesses a nonvanishing vorticity arising from angular variations,
\begin{equation}
\omega_z^{\rm near}
=
\frac{\sigma}{2\pi\eta_s}
\frac{\sin\!\big[2(\alpha-\theta)\big]}{r^2}.
\end{equation}
Thus, the near-field flow is divergence-free, quadrupolar in angle, and decays as $v\sim r^{-1}$, $\omega\sim r^{-2}$.

In the far field ($r\gg\lambda$), hydrodynamic screening modifies both the magnitude and structure of the flow. The velocity acquires an azimuthal component,
\begin{align}
v_r^{\rm far}
&=
-\frac{\lambda\sigma}{2\pi\eta_s}
\left(\frac{1}{r^2}-\frac{2\lambda}{r^3}\right)
\cos\!\big[2(\alpha-\theta)\big],
\\
v_\theta^{\rm far}
&=
\frac{\lambda\sigma}{4\pi\eta_s}
\left(\frac{1}{r^2}-\frac{4\lambda}{r^3}\right)
\sin\!\big[2(\alpha-\theta)\big],
\end{align}
leading to a slower decay of vorticity,
\begin{equation}
\omega_z^{\rm far}
=
\frac{3\lambda\sigma}{4\pi\eta_s}
\frac{\sin\!\big[2(\alpha-\theta)\big]}{r^3}.
\end{equation}
All contributions of order $r^{-4}$ cancel identically, and the numerical factor $3$ originates from the unequal prefactors of the radial and azimuthal velocity components. In this regime, both velocity components scale as $r^{-2}$, while $\omega\sim r^{-3}$.

Since the membrane flow is incompressible, $\nabla\!\cdot\!\mathbf v=0$, the velocity admits a streamfunction representation with
\(
v_r=(1/r)\partial_\theta\Psi,\;
v_\theta=-\partial_r\Psi.
\)
The corresponding streamfunctions are
\begin{equation}
\Psi^{\rm near}
=
\frac{\sigma}{8\pi\eta_s}
\sin\!\big[2(\alpha-\theta)\big],
\qquad
\Psi^{\rm far}
=
\frac{\lambda\sigma}{4\pi\eta_s}
\left(\frac{1}{r}-\frac{2\lambda}{r^2}\right)
\sin\!\big[2(\alpha-\theta)\big],
\end{equation}
both exhibiting a quadrupolar angular structure. The distinct radial dependence reflects the crossover from an unscreened near-field flow to a screened far-field regime controlled by $\lambda$. These results provide the fundamental building blocks for the interaction and dynamics of multiple dipoles in viscous membranes.

\section{Two-body Hamiltonian structure in the quenched limit}
\label{sec:two_body_ham}
\begin{figure}[t]
    \centering
    \includegraphics[width=\linewidth]{images/phase.png}
    \caption{
    Phase portraits of the quenched two-dipole dynamics in the $(r,\psi)$ plane for pullers ($\sigma>0$). 
    In the near field (left), $\psi$ is conserved ($\dot\psi=0$), and trajectories remain on horizontal lines; the dashed lines at $\psi=\pm\pi/4$ mark the null directions separating aggregating and separating sectors. 
    In the far field (right), radial and angular dynamics are coupled, and generic trajectories are driven toward the aligned manifold $\psi=0$, leading to aggregation. 
    The figure illustrates the qualitative reorganization of Hamiltonian phase space across the Saffman crossover.
    }
    \label{fig:phase_portraits}
\end{figure}
We now consider the interaction of two force dipoles in the quenched-orientation limit, where the dipole angles \(\alpha_1\) and \(\alpha_2\) are fixed. In the near zone, the single-dipole flow is purely radial, and the relative coordinate \(\mathbf R_{12}=\mathbf R_1-\mathbf R_2\), with \(r_{12}=|\mathbf R_{12}|\) and \(\theta_{12}=\arg(\mathbf R_{12})\), obeys
\begin{equation}
\dot r_{12}
=
-\frac{1}{4\pi\eta_s r_{12}}
\left[
\sigma_2 \cos\!\big(2(\alpha_2-\theta_{12})\big)
+
\sigma_1 \cos\!\big(2(\alpha_1-\theta_{12})\big)
\right],
\qquad
\dot\theta_{12}=0.
\end{equation}
Thus the line of centers is conserved, and the two-body dynamics reduces to a single scalar degree of freedom. Introducing canonical variables
\begin{equation}
q=\theta_{12},
\qquad
p=-\frac{r_{12}^2}{2},
\end{equation}
one obtains
\begin{equation}
\dot q=\frac{\partial H}{\partial p}=0,
\qquad
\dot p=-\frac{\partial H}{\partial q},
\end{equation}
generated by the Hamiltonian
\begin{equation}
H(q)
=
\frac{1}{8\pi\eta_s}
\left[
\sigma_2 \sin\!\big(2(\alpha_2-q)\big)
+
\sigma_1 \sin\!\big(2(\alpha_1-q)\big)
\right].
\end{equation}
Because \(H\) is independent of \(p\), the angular variable is frozen, and the radial motion integrates exactly as
\begin{equation}
r_{12}^2(t)=r_{12}^2(0)-2C(q)\,t,
\qquad
C(q)=\frac{1}{4\pi\eta_s}
\left[
\sigma_2 \cos\!\big(2(\alpha_2-q)\big)
+
\sigma_1 \cos\!\big(2(\alpha_1-q)\big)
\right].
\end{equation}
The sign of \(C(q)\) determines whether the dipoles separate or collapse in finite time.

For the symmetric coaligned case,
\begin{equation}
\alpha_1=\alpha_2=\alpha,
\qquad
\sigma_1=\sigma_2=\sigma,
\end{equation}
the Hamiltonian reduces to
\begin{equation}
H(q)=\frac{\sigma}{4\pi\eta_s}\sin\!\big(2(\alpha-q)\big),
\end{equation}
and the dynamics is governed by the constant mismatch angle \(\psi=\alpha-\theta_{12}\),
\begin{equation}
\dot r_{12}
=
-\frac{\sigma}{2\pi\eta_s}\frac{\cos(2\psi)}{r_{12}},
\qquad
r_{12}^2(t)=r_{12}^2(0)-\frac{\sigma}{\pi\eta_s}\cos(2\psi)\,t.
\end{equation}
Hence \(\cos 2\psi>0\) gives attraction and finite-time collapse, \(\cos 2\psi<0\) gives separation, and \(\cos 2\psi=0\) gives constant separation. In particular, for \(\psi=0\),
\begin{equation}
r_{12}^2(t)=r_{12}^2(0)-\frac{\sigma}{\pi\eta_s}t,
\qquad
t_c=\frac{\pi\eta_s r_{12}^2(0)}{\sigma},
\end{equation}
corresponding to maximal aggregation.

In the far zone, the screened single-dipole flow contains both radial and azimuthal components, and the leading relative dynamics becomes
\begin{align}
\dot r_{12}
&=
-\frac{\lambda}{2\pi\eta_s}
\left(\frac{1}{r_{12}^2}\right)
\left[
\sigma_2\cos\!\big(2(\alpha_2-\theta_{12})\big)
+
\sigma_1\cos\!\big(2(\alpha_1-\theta_{12})\big)
\right],
\\[4pt]
\dot\theta_{12}
&=
\frac{\lambda}{4\pi\eta_s}
\left(\frac{1}{r_{12}^3}\right)
\left[
\sigma_2\sin\!\big(2(\alpha_2-\theta_{12})\big)
+
\sigma_1\sin\!\big(2(\alpha_1-\theta_{12})\big)
\right].
\end{align}
Unlike the near field, the angular coordinate now evolves nontrivially. Nevertheless, the problem still admits an exact one-degree-of-freedom Hamiltonian reduction in the same canonical variables,
\begin{equation}
H(q,p)
=
\frac{\lambda}{4\pi\eta_s}
\left(\frac{1}{r_{12}}\right)
\left[
\sigma_2\sin\!\big(2(\alpha_2-q)\big)
+
\sigma_1\sin\!\big(2(\alpha_1-q)\big)
\right],
\qquad
r_{12}=\sqrt{-2p},
\end{equation}
with \(\dot q=\partial H/\partial p\) and \(\dot p=-\partial H/\partial q\). In the symmetric coaligned case this becomes
\begin{equation}
H(q,p)
=
\frac{\lambda\sigma}{2\pi\eta_s}
\left(\frac{1}{r_{12}}\right)
\sin\!\big(2(\alpha-q)\big),
\end{equation}
and, writing \(\psi=\alpha-\theta_{12}\), the reduced equations as before
\begin{align}
\dot r_{12}
&=
-\frac{\lambda\sigma}{\pi\eta_s}
\left(\frac{1}{r_{12}^2}\right)\cos(2\psi),
\\[4pt]
\dot\psi
&=
-\frac{\lambda\sigma}{2\pi\eta_s}
\left(\frac{1}{r_{12}^3}\right)\sin(2\psi).
\end{align}
The far-zone problem therefore remains Hamiltonian but becomes intrinsically two-dimensional: the relative separation and angular mismatch are coupled, and the invariant directions \(\psi=0\) and \(\psi=\pi/2\) replace the frozen-angle structure of the near field. For \(\psi=0\), the angular dynamics vanishes and the radial evolution reduces to
\begin{equation}
\dot r_{12}
=
-\frac{\lambda\sigma}{\pi\eta_s}
\left(\frac{1}{r_{12}^2}\right),
\end{equation}
providing the screened far-field analogue of aligned near-zone collapse. Thus the Saffman crossover reorganizes not only the decay of the underlying dipolar flow, but also the Hamiltonian phase-space structure of the two-body problem: the near field is purely radial with frozen angular dynamics, whereas the far field exhibits genuine radial-angular coupling. 
The qualitative structure of the two-body dynamics is summarized in the phase portraits shown in Fig.~\ref{fig:phase_portraits}. In the near field, the Hamiltonian is independent of the conjugate momentum, and the angular mismatch angle $\psi$ is conserved. As a result, the dynamics is effectively one-dimensional: trajectories remain on horizontal lines in the $(r,\psi)$ plane, and the sign of $\cos(2\psi)$ determines whether the pair undergoes aggregation or separation. The null directions $\psi=\pm\pi/4$ act as separatrices between these regimes.

In contrast, the far-field dynamics is intrinsically two-dimensional. Hydrodynamic screening introduces a dependence of the Hamiltonian on the radial coordinate, leading to coupled evolution of $r$ and $\psi$. As seen in Fig.~\ref{fig:phase_portraits}, generic trajectories are driven toward the aligned manifold $\psi=0$, which acts as an attractor for pullers ($\sigma>0$). Once the system enters the sector $|\psi|<\pi/4$, the radial dynamics becomes strictly attractive, resulting in monotonic contraction of the separation. Thus, except for a measure-zero set of initial conditions lying on the null directions, the far-field dynamics generically leads to aggregation.

This comparison highlights a fundamental reorganization of the Hamiltonian phase space across the Saffman crossover: the near field exhibits a degenerate, effectively one-dimensional flow with frozen angular dynamics, whereas the far field supports a fully coupled phase-space structure with an emergent attracting manifold that controls the asymptotic aggregation dynamics. A complementary view of this reorganization is provided by the collapse laws implied by the Hamiltonian dynamics. In the near field, the independence of the Hamiltonian from the conjugate momentum freezes the angular degree of freedom, yielding purely radial evolution,
\(
r^2(t)=r_0^2-\frac{\sigma}{\pi\eta_s}\cos(2\psi)\,t,
\)
so that aggregation proceeds at a rate set by the initial mismatch $\psi$. In contrast, in the far field the coupling between $r$ and $\psi$ drives the system toward the aligned manifold $\psi\to 0$, leading to universal late-time behavior. Using the first integral $R=R_0\,\sin(2\psi)/\sin(2\psi_0)$ together with the implicit evolution, one finds near collapse
\(
\psi\sim (t_c-t)^{1/3},
\qquad
R^3(t)\sim \frac{3\sigma\lambda}{\pi\eta_s}(t_c-t),
\)
independent of initial conditions in the asymptotic limit $t\to t_c$. Thus, while the near-field dynamics retains memory of the initial angular mismatch, the far-field dynamics exhibits a universal cubic collapse law governed by an attracting manifold. This provides a direct dynamical signature of the Saffman crossover: a transition from degenerate Hamiltonian flow with frozen angles to a fully coupled phase-space structure in which angular dynamics self-organizes aggregation.
\section{Summary and conclusion}
\label{cncl}

We have developed an analytic description of force-dipole hydrodynamics in a viscous membrane coupled to an infinite surrounding fluid and shown that the Saffman length
\(
\lambda=\eta_s/(\eta_++\eta_-)
\)
controls a sharp crossover in both the structure of the dipolar flow and the effective two-body dynamics of quenched dipoles. Starting from the membrane Green's function, we obtained the dipolar velocity field by differentiation along the dipole axis and identified two asymptotic regimes. In the near zone \(r\ll\lambda\), the flow reduces to an effectively two-dimensional form,
\[
\mathbf v_{\rm near}(\mathbf r)
=
\frac{\sigma}{4\pi\eta_s}\frac{1-2\gamma^2}{r}\,\hat{\mathbf r},
\qquad
\omega_{\rm near}(\mathbf r)
=
\frac{\sigma}{\pi\eta_s}\frac{\gamma\beta}{r^2},
\]
so that the leading dipolar flow is purely radial, with quadrupolar angular dependence and scalings \(v\sim r^{-1}\), \(\omega\sim r^{-2}\). In the far zone \(r\gg\lambda\), momentum leakage into the bulk screens the interaction and produces a qualitatively different flow,
\[
\mathbf v_{\rm far}(\mathbf r)
=
\frac{\sigma\lambda}{2\pi\eta_s r^2}
\left[
\gamma\,\hat{\mathbf d}
+
(1-3\gamma^2)\hat{\mathbf r}
\right]
+
\mathcal O(r^{-3}),
\qquad
\omega_{\rm far}(\mathbf r)
=
\frac{3\sigma\lambda}{2\pi\eta_s}\frac{\gamma\beta}{r^3},
\]
which is no longer purely radial and exhibits the screened scalings \(v\sim r^{-2}\), \(\omega\sim r^{-3}\).

For two identical quenched and coaligned dipoles, this crossover reorganizes the relative dynamics in a striking way. In the near zone, the pair dynamics is exactly solvable and effectively one-dimensional:
\[
\dot R
=
-\frac{\sigma}{2\pi\eta_s}\frac{\cos 2\psi}{R},
\qquad
R^2(t)
=
R_0^2
-
\frac{\sigma}{\pi\eta_s}\cos 2\psi\,(t-t_0),
\]
with \(\psi\) the fixed angle between the dipole axis and the line of centers. The relative trajectory remains strictly collinear, and for pullers (\(\sigma>0\)) the aligned branch \(\psi=0\) undergoes finite-time collapse with \(R\sim (t_c-t)^{1/2}\). In contrast, the far-zone problem remains integrable but becomes intrinsically two-dimensional:
\[
\dot R
=
-\frac{\sigma\lambda}{\pi\eta_s}\frac{\cos 2\psi}{R^2},
\qquad
\dot\psi
=
-\frac{\sigma\lambda}{2\pi\eta_s}\frac{\sin 2\psi}{R^3}.
\]
The dynamics now couples separation and orientation, admits the first integral
\[
R
=
R_0\frac{\sin 2\psi}{\sin 2\psi_0},
\]
and drives pullers toward the stable aligned manifold \(\psi\to 0\) (while for pushers the stable manifold is \(\psi\to \pi/2\)). As a result, the late-time aggregation becomes universal:
\[
R^3(t)\sim \frac{3\sigma\lambda}{\pi\eta_s}(t_c-t),
\]
independent of the initial mismatch angle. Thus the far-field collapse law is cubic, \(R\sim (t_c-t)^{1/3}\), rather than the square-root law of the near field.

The quenched two-body problem also admits a compact Hamiltonian formulation in both asymptotic regimes. In the near field, the Hamiltonian is independent of the conjugate radial momentum, which freezes the angular degree of freedom and explains the effectively one-dimensional character of the motion. In the far field, the Hamiltonian acquires an explicit radial dependence through the screened interaction, leading to genuine radial-angular coupling and an emergent attracting manifold that controls asymptotic aggregation. The Saffman crossover therefore reorganizes not only the spatial decay of membrane dipolar flows, but also the phase-space structure of the corresponding Hamiltonian dynamics.

These results provide a minimal analytic framework for understanding membrane-mediated interactions of active force dipoles with fixed orientation. They show that hydrodynamic screening in membranes has consequences beyond a simple modification of decay exponents: it changes the geometry of the flow, the nature of the two-body trajectories, and the universality class of finite-time collapse. Natural extensions include orientable dipoles with internal rotational dynamics, interactions among many dipoles and dipolar clusters, and the role of confinement, asymmetry of the surrounding fluids, or elastic membrane effects in modifying the crossover structure.
\section{Acknowledgments}
It is a pleasure to thank Sarthak Bagaria, Michael D Graham, Mark Henle, Naomi Oppenheimer and Haim Diamant.
R.S.\ acknowledges support from the DST INSPIRE Faculty Fellowship, India (Grant No.\ IFA19-PH231), as well as the NFSG and the OPERA Research Grant from the Birla Institute of Technology and Science, Pilani (Hyderabad Campus). 
\section{Data Availability}
The data that support the findings of this study are available within the article. 

\appendix
\section{Recap on Green's function of a free viscous membrane coupled to an infinite subphase and Fourier inversion to real space}
\label{app1}
In this section, we derive the hydrodynamic Green’s function for an incompressible two-dimensional viscous membrane embedded in a three-dimensional viscous fluid extending to infinity on both sides. This setup corresponds to the “free membrane” geometry originally analyzed by Saffman and Delbrück and later generalized by Hughes, Pailthorpe, and White.

Although this material is standard, we have not found a source that presents the complete derivation in one place, particularly the Fourier-inversion. We therefore provide a self-contained account, including detailed intermediate integral identities obtained using \textit{Mathematica}.\\\\
We consider an incompressible membrane with two-dimensional viscosity
$\eta_{2D}$, located at $z=0$, surrounded by identical bulk fluids of viscosity
$\eta$ in the half-spaces $z>0$ and $z<0$.
The membrane velocity field $\mathbf{v}(\mathbf{r})$ satisfies
\begin{equation}
\nabla \cdot \mathbf{v} = 0 ,
\end{equation}
and the in-plane Stokes equation
\begin{equation}
-\nabla p
+ \eta_{2D} \nabla^2 \mathbf{v}
+ \mathbf{f}
+ \mathbf{t}^{(+)} + \mathbf{t}^{(-)} = 0 ,
\label{eq:membraneStokes}
\end{equation}
where $\mathbf{f}(\mathbf{r})$ is an applied in-plane force density and
$\mathbf{t}^{(\pm)}$ are the tangential traction forces exerted by the bulk fluids
above and below the membrane. The surrounding fluids obey the three-dimensional Stokes equations
\begin{equation}
\eta \nabla^2 \mathbf{u}^{(\pm)} - \nabla P^{(\pm)} = 0,
\qquad
\nabla \cdot \mathbf{u}^{(\pm)} = 0 ,
\end{equation}
with the no-slip boundary condition
\begin{equation}
\mathbf{u}^{(\pm)}(\mathbf{r},z=0) = \mathbf{v}(\mathbf{r}).
\end{equation}
We first make explicit how the three-dimensional bulk flow produces a
nonlocal traction term in the membrane Stokes equation.
Working in Fourier space with respect to the in-plane coordinates,
we write
\begin{equation}
\mathbf{v}(\mathbf{r})
=
\int \frac{d^2 q}{(2\pi)^2}\,
\mathbf{v}(\mathbf{q})\,e^{i\mathbf{q}\cdot\mathbf{r}},
\qquad
\mathbf{u}^{(\pm)}(\mathbf{r},z)
=
\int \frac{d^2 q}{(2\pi)^2}\,
\mathbf{u}^{(\pm)}(\mathbf{q},z)\,e^{i\mathbf{q}\cdot\mathbf{r}} .\nn
\end{equation}
For a given in-plane wavevector $\mathbf{q}$, the bulk Stokes equations reduce to
\begin{equation}
\eta(\partial_z^2 - q^2)\mathbf{u}^{(\pm)} - i\mathbf{q} P^{(\pm)} = 0,
\qquad
i\mathbf{q}\cdot\mathbf{u}^{(\pm)} + \partial_z u_z^{(\pm)} = 0 .
\end{equation}
Regularity as $|z|\to\infty$ implies
\begin{equation}
\mathbf{u}^{(\pm)}(\mathbf{q},z)
=
\mathbf{v}(\mathbf{q})\,e^{-q|z|}.\nn
\end{equation}
The tangential traction exerted by the bulk fluid on the membrane is
\begin{equation}
\mathbf{t}^{(\pm)}(\mathbf{q})
=
\pm \eta\,\partial_z \mathbf{u}^{(\pm)}\big|_{z=0^\pm}
=
- \eta q\,\mathbf{v}(\mathbf{q}).\nn
\end{equation}
Since the subphases above and below the membrane are identical,
the total traction is
\begin{equation}
\mathbf{t}^{(+)} + \mathbf{t}^{(-)}
=
-2\eta q\,\mathbf{v}(\mathbf{q}).\nn
\end{equation}
The Fourier-transformed membrane Stokes equation becomes
\begin{equation}
-\mathrm{i}q_i p(\mathbf{q})
+
\left(\eta_{2D}q^2 + 2\eta q\right)v_i(\mathbf{q})
=
f_i(\mathbf{q}).
\label{eq:FourierStokes}
\end{equation}
Membrane incompressibility,
\begin{equation}
q_i v_i(\mathbf{q})=0,\nn
\end{equation}
allows the pressure to be eliminated using the transverse projection operator
\begin{equation}
\mathcal{P}_{ij}(\mathbf{q})
=
\delta_{ij}-\frac{q_i q_j}{q^2}.\nn
\end{equation}
Applying $\mathcal{P}_{ij}$ to Eq.~\eqref{eq:FourierStokes} yields
\begin{equation}
v_i(\mathbf{q})
=
\frac{\delta_{ij}-q_i q_j/q^2}
{\eta_{2D}q^2+2\eta q}\,f_j(\mathbf{q})
=
\frac{\delta_{ij}-q_i q_j/q^2}
{\eta_{2D}q\left(q+\lambda^{-1}\right)}\,f_j(\mathbf{q}),
\end{equation}
where $\lambda=\eta_{2D}/(2\eta)$. The real-space Green's tensor is defined by the inverse Fourier transform
\begin{equation}
G_{ij}(\mathbf r)
=
\int_{\mathbb R^2}\frac{d^2q}{(2\pi)^2}\,
e^{i\mathbf q\cdot\mathbf r}\,G^{\rm free}_{ij}(\mathbf q).
\label{eq:Gij_invFT_def}
\end{equation}
Throughout, $q\equiv|\mathbf q|$ and $r\equiv|\mathbf r|$. Substituting the Green's function gives
\begin{equation}
G_{ij}(\mathbf r)
=
\frac{1}{\eta_{2D}}
\int\frac{d^2q}{(2\pi)^2}\,
e^{i\mathbf q\cdot\mathbf r}\,
\frac{\delta_{ij}-q_i q_j/q^2}{q\,(q+\lambda^{-1})}.
\label{eq:Gij_invFT_start}
\end{equation}
Choosing coordinates such that $\mathbf r=r\,\hat{\mathbf x}$ and introducing  polar coordinates for $\mathbf q$:
\begin{equation}
\mathbf q = q(\cos\theta,\sin\theta),\qquad
d^2q = q\,dq\,d\theta,\qquad
\mathbf q\cdot\mathbf r = q r\cos\theta.\nn
\label{eq:q_polar}
\end{equation}
In these variables,
\begin{equation}
\frac{q_i q_j}{q^2}
=
\begin{pmatrix}
\cos^2\theta & \sin\theta\cos\theta\\
\sin\theta\cos\theta & \sin^2\theta
\end{pmatrix},
\qquad
\delta_{ij}-\frac{q_i q_j}{q^2}
=
\begin{pmatrix}
\sin^2\theta & -\sin\theta\cos\theta\\
-\sin\theta\cos\theta & \cos^2\theta
\end{pmatrix}.\nn
\label{eq:projector_components}
\end{equation}
By symmetry under reflection about the $x$-axis ($\theta \mapsto 2\pi - \theta$), the off-diagonal terms, proportional to $\sin\theta\cos\theta$, are odd, while the remaining factors are even. Their angular integral therefore vanishes, yielding
\begin{equation}
G_{xy}(\mathbf r)=G_{yx}(\mathbf r)=0 
\qquad \text{for}\quad \mathbf r = r\hat{\mathbf x}.\nn
\label{eq:Gxy_zero}
\end{equation}
Using $d^2q=q\,dq\,d\theta$, the factor $q$ in the measure cancels the
explicit $1/q$ in \eqref{eq:Gij_invFT_start}, leaving the kernel $dq/(q+\lambda^{-1})$.
Thus,
\begin{align}
G_{xx}(r)
&=
\frac{1}{\eta_{2D}}
\int_0^\infty\frac{dq}{(2\pi)^2}\,
\frac{1}{q+\lambda^{-1}}
\int_0^{2\pi} d\theta\,
e^{iqr\cos\theta}\,\sin^2\theta,\nn
\label{eq:Gxx_theta_int}
\\
G_{yy}(r)
&=
\frac{1}{\eta_{2D}}
\int_0^\infty\frac{dq}{(2\pi)^2}\,
\frac{1}{q+\lambda^{-1}}
\int_0^{2\pi} d\theta\,
e^{iqr\cos\theta}\,\cos^2\theta.\nn
\end{align}
We now evaluate the two angular integrals explicitly.
First we use the trigonometric identities
\begin{equation}
\sin^2\theta=\frac12\left(1-\cos2\theta\right),\qquad
\cos^2\theta=\frac12\left(1+\cos2\theta\right),
\label{eq:sin2_cos2_halves}
\end{equation}
so that
\begin{align}
\int_0^{2\pi} d\theta\,e^{iqr\cos\theta}\sin^2\theta
&=
\frac12\int_0^{2\pi} d\theta\,e^{iqr\cos\theta}
-\frac12\int_0^{2\pi} d\theta\,e^{iqr\cos\theta}\cos2\theta,
\label{eq:ang_sin2_split}
\\
\int_0^{2\pi} d\theta\,e^{iqr\cos\theta}\cos^2\theta
&=
\frac12\int_0^{2\pi} d\theta\,e^{iqr\cos\theta}
+\frac12\int_0^{2\pi} d\theta\,e^{iqr\cos\theta}\cos2\theta.
\label{eq:ang_cos2_split}
\end{align}
Next, we use the standard Bessel representations (for real $z$)
\begin{equation}
\int_0^{2\pi} d\theta\,e^{iz\cos\theta}=2\pi J_0(z),
\qquad
\int_0^{2\pi} d\theta\,e^{iz\cos\theta}\cos(2\theta)=-2\pi J_2(z).
\label{eq:bessel_angular_ids}
\end{equation}
Substituting \eqref{eq:bessel_angular_ids} into \eqref{eq:ang_sin2_split}-\eqref{eq:ang_cos2_split}
yields
\begin{align}
\int_0^{2\pi} d\theta\,e^{iqr\cos\theta}\sin^2\theta
&=
\frac12(2\pi J_0(qr))-\frac12(-2\pi J_2(qr))
=
\pi\big[J_0(qr)+J_2(qr)\big],
\label{eq:ang_sin2_final}
\\
\int_0^{2\pi} d\theta\,e^{iqr\cos\theta}\cos^2\theta
&=
\frac12(2\pi J_0(qr))+\frac12(-2\pi J_2(qr))
=
\pi\big[J_0(qr)-J_2(qr)\big].
\label{eq:ang_cos2_final}
\end{align}
and hence we obtain 
\begin{align}
G_{xx}(r)
&=
\frac{1}{4\pi\eta_{2D}}
\int_0^\infty dq\,
\frac{J_0(qr)+J_2(qr)}{q+\lambda^{-1}},
\label{eq:Gxx_radial}
\\
G_{yy}(r)
&=
\frac{1}{4\pi\eta_{2D}}
\int_0^\infty dq\,
\frac{J_0(qr)-J_2(qr)}{q+\lambda^{-1}}.
\label{eq:Gyy_radial}
\end{align}
Introducing the dimensionless variable
\begin{equation}
x \equiv \frac{r}{\lambda}\nn
\end{equation}
the Fourier inversion reduces to evaluating two Hankel-type integrals. We will use the following identities (verified in Mathematica ):
\begin{equation}
\int_0^\infty \frac{J_0(qr)}{q+\lambda^{-1}}\,dq
=
\frac{\pi}{2}\Big[H_0(x)-Y_0(x)\Big].
\label{eq:J0_identity_verified}
\end{equation}
and
\begin{equation}
\int_0^\infty \frac{J_2(qr)}{q+\lambda^{-1}}\,dq
=
\lambda\left[
\frac{\pi}{2}\Big(H_2(x)-Y_2(x)\Big)
-\frac{2}{x^2}
-\frac{x}{3}
\right],
\qquad x=\frac{r}{\lambda}.
\label{eq:J2_identity_dimensional_verified}
\end{equation}
Finally, using Struve relation
\begin{equation}
H_2(x)=\frac{2}{x}H_1(x)-H_0(x)+\frac{2x}{3\pi}.
\label{strvid}
\end{equation}
we arrive at 
\begin{align}
G_{xx}(r)
&=
\frac{1}{4\pi\eta_{2D}}\Big[I_0(r)+I_2(r)\Big],\nn
\\
G_{yy}(r)
&=
\frac{1}{4\pi\eta_{2D}}\Big[I_0(r)-I_2(r)\Big].
\label{eq:Gyy_I0I2}
\end{align}
where \begin{align}
I_0(r) &= \frac{\pi}{2}\Big[H_0(x)-Y_0(x)\Big],\nn
\\
I_2(r) &= \left[
\frac{\pi}{2}\Big(H_2(x)-Y_2(x)\Big)
-\frac{2}{x^2}
-\frac{x}{3}
\right].\nn
\end{align}
By rotational invariance in the plane, $G_{ij}(\mathbf r)$ must have the form
\begin{equation}
G_{ij}(\mathbf r)
=
\frac{1}{4\pi\eta_{2D}}
\left[
A_0(x)\,\delta_{ij}
+
A_1(x)\,\frac{r_i r_j}{r^2}
\right],
\qquad x=\frac{r}{\lambda}.
\label{eq:tensor_decomp_AB}
\end{equation}
With $\mathbf r=r\hat{\mathbf x}$, we have $r_i r_j/r^2=\mathrm{diag}(1,0)$ and therefore
\begin{equation}
G_{yy}(r)=\frac{1}{4\pi\eta_{2D}}A_0(x),
\qquad
G_{xx}(r)=\frac{1}{4\pi\eta_{2D}}\big[A_0(x)+A_1(x)\big].\nn
\label{eq:GxxGyy_AB_relations}
\end{equation}
which implies
\begin{equation}
A_0(x)=4\pi\eta_{2D}\,G_{yy}(r),
\qquad
A_1(x)=4\pi\eta_{2D}\,\big[G_{xx}(r)-G_{yy}(r)\big].
\label{eq:A_B_from_components}
\end{equation}
Thus we find
\begin{equation}
A_0(x)
=
\frac{\pi}{2}\Big[H_0(x)-Y_0(x)-H_2(x)+Y_2(x)\Big]
+\frac{2}{x^2}
+\frac{x}{3},\nn
\end{equation}
\begin{equation}
A_1(x)
=
\pi\Big[H_2(x)-Y_2(x)\Big]
-\frac{4}{x^2}
-\frac{2x}{3}.\nn
\end{equation}
Rewriting $A$ and $B$ in terms of $H_0,H_1$ only, the $x/3$ term in $A(x)$ cancels. Inserting the Struve identity \eqref{strvid} we get
\begin{equation}
A_1(x)=
-\pi H_0(x)+\frac{2\pi}{x}H_1(x)-\pi Y_2(x)-\frac{4}{x^2}.
\label{eq:B_noH2_final}
\end{equation}
and
\begin{equation}
A_0(x)
=
\pi H_0(x)-\frac{\pi}{x}H_1(x)
+\frac{2}{x^2}
-\frac{\pi}{2}\Big[Y_0(x)-Y_2(x)\Big].
\label{eq:A_noH2_final}
\end{equation}
as reported in the main text, 
together with the final
real-space Green's tensor:
\begin{equation}
G_{ij}(\mathbf r)
=
\frac{1}{4\pi\eta_{2D}}
\left[
A\!\left(\frac{r}{\lambda}\right)\delta_{ij}
+
B\!\left(\frac{r}{\lambda}\right)\frac{r_i r_j}{r^2}
\right].
\label{eq:Gij_final_boxed}
\end{equation}
where $
x \equiv \frac{r}{\lambda}.
$ 
\begin{footnotesize}

\end{footnotesize}


\begin{thebibliography}{99}
\bibitem{purcell}E.M. Purcell, \textit{Life at low Reynolds number}, American Journal of Physics 45, 3 (1977)
\bibitem{lauga2009} E. Lauga and T.R. Powers,\textit{The hydrodynamics of swimming microorganisms}, Rep. Prog. Phys. \textbf{72}, 096601 (2009)

\bibitem{lamb} H.~Lamb, \textit{Hydrodynamics} (University Press, 1924)
\bibitem{saff1}P.~G.~Saffman,   \textit{Brownian motion in thin sheets of viscous fluid}, J. Fluid Mech. 73:593–602, (1975)
\bibitem{saff2} P.~G.~Saffman and M.~Delbr{\"u}ck,\textit{ Brownian motion in biological
membranes},  Proc. Natl. Acad. Sci. USA. 72:3111–3113 (1975)
\bibitem{hughes} B.~D.~Hughes, B.~A.~Pailthorpe, and L.~R.~White,\textit{ The translational and rotational drag on a cylinder moving in a membrane},
J. Fluid Mech. \textbf{110}, 349–372 (1981)
\bibitem {evans} E.~Evans, and E.~Sackmann, \textit{Translational and rotational drag coefficients for a disk moving in a liquid membrane associated with a rigid substrate}, J. Fluid Mech. \textbf{194}, 553–561 (1988)

\bibitem{lg96} D.K. Lubensky and R.E. Goldstein, \textit{Hydrodynamics of monolayer domains at the
air–water interface}, Physics of Fluids \textbf{8}, 843 (1996)
\bibitem{staj} H.~A.~Stone, and A.~Ajdari. \textit{Hydrodynamics of particles embedded in a flat surfactant layer overlying a subphase of finite depth}, Journal of Fluid Mechanics \textbf{369} (1998)
\bibitem{fischer} T.M. Fischer, \textit{The drag on needles moving in a Langmuir
monolayer}, J. Fluid. Mech. \textbf{498}, 123 (2004)
\bibitem{OppenheimerDiamant2009}
N.~Oppenheimer and H.~Diamant,
\emph{Correlated diffusion of membrane proteins and their effect on membrane viscosity},
Biophys.\ J.\ \textbf{96}, 3041-3049 (2009).

\bibitem{OppenheimerDiamant2010}
N.~Oppenheimer and H.~Diamant,
\emph{Correlated dynamics of inclusions in a supported membrane},
Phys.\ Rev.\ E \textbf{82}, 041912 (2010).

\bibitem{OppenheimerDiamant2011}
N.~Oppenheimer and H.~Diamant,
\emph{Dynamics of membranes with immobile inclusions},
Phys.\ Rev.\ Lett.\ \textbf{107}, 258102 (2011).


\bibitem{mik} Alexander S Mikhailov and Raymond Kapral, \textit{Hydrodynamic Collective Effects of Active
Protein Machines in Solution and Lipid Bilayers}, Proceedings of the National Academy of Sciences, 112 (\textbf{28}): E3639 (2015)
\bibitem{mnk}  H. Manikantan, \textit{Tunable Collective Dynamics of Active Inclusions in Viscous
Membranes}, Phys. Rev. Letters \textbf{125}, 268101 (2020)
\bibitem{Graaf}  J.D. Graaf, J. Stenhammar, \textit{Stirring by Periodic Arrays of Microswimmers}, J. Fluid Mech. \textbf{811}, 487-498 (2017)

\bibitem{aggexp1}O. Campas, C. Leduc, P. Bassereau, J. Casademunt, J.F. Joanny, J. Prost, \textit{Coordination of Kinesin Motors Pulling on Fluid Membranes}, Biophysical Journal, \textbf{94}, 5009 (2008)
\bibitem{aggexp2} R. Grover, J. Fischer, F.W. Schwarz, W.J. Walter, P. Schwille and S. Diez, \textit{Transport efficiency of membrane-anchored kinesin-1 motors depends on motor density and diffusivity}, PNAS \textbf{113}, E7185-E7193 (2016)
\bibitem{BagariaSamanta2022}
S.~Bagaria and R.~Samanta,
\emph{Dynamics of force dipoles in curved fluid membranes},
Phys.\ Rev.\ Fluids \textbf{7}, 093101 (2022).


\bibitem {henlev2008} M.~L.~Henle, R.~McGorty, A.~B.~Schofield, A.~D.~Dinsmore, A.~J.~Levine, \textit {The effect of curvature and topology on membrane hydrodynamics}, EPL (Europhysics Letters), \textbf{84}, 48001 (2008)
\bibitem{henlev2010} M.~L~Henle, A.~J.~Levine, \textit {Hydrodynamics in curved membranes: The effect of geometry on particulate mobility}, Physical Review E, \textbf{81}, 011905 (2010)
\bibitem{wg2012} F.~G.~Woodhouse and R.~E.~Goldstein, \textit {Shear-Driven Circulation Patterns in Lipid Membrane Vesicles},  J. Fluid Mech. \textbf{705}, 165 (2012)
\bibitem{wg2013} A.~R.~H.~Smith, F.~G.~Woodhouse, V.~Kantsler, R.~E.~Goldstein, \textit{ Membrane Viscosity Determined from Shear-Driven Flow in Giant Vesicles}, PRL \textbf{111}, 038103 (2013)
\bibitem {atzbergershape} F.~Quemeneura, J.~K.~Sigurdsson, M.~Rennerf, P.~J.~Atzberger, P.~Bassereaua, and D.~Lacoste, \textit{ Shape matters in protein mobility within membranes}, Proceedings of the National Academy of Sciences (PNAS),\textbf{ 11}, No. 14, pg. 5083–5087, (2014)
\bibitem {atz2016}  J.~K.~Sigurdsson and P.~J.~Atzberger,  \textit{Hydrodynamic Coupling of Particle Inclusions Embedded in Curved Lipid Bilayer Membranes}, \textbf{12}, 6685-6707, Soft Matter, The Royal Society of Chemistry, (2016)
\bibitem{atz2018} B.~J.~Gross and P.~J.~Atzberger, \textit{Hydrodynamic Flows on Curved Surfaces : Spectral Numerical Methods for radial manifold shapes}, Journal of Computational Physics, \textbf{371}, 663 (2018)
\bibitem{atz2019} D.~Rower, M.~Padidar and P.~J.~Atzberger, \textit{Surface Fluctuating Hydrodynamics methods for the drift-diffusion dynamics of particles and microstructures within curved fluid interfaces}, arXiv 1906.01146
\bibitem{JainSamanta2023}
S.~Jain and R.~Samanta,
\emph{Force dipole interactions in tubular fluid membranes},
Phys.\ Fluids \textbf{35}, 071901 (2023).
\bibitem{KrishnanOddViscosity2026}
S.~Krishnan, U.~Maurya, and R.~Samanta,
\textit{Force Dipole Interactions in Membranes with Odd Viscosity},
arXiv:2603.07830 [cond-mat.soft], 2026.
\bibitem{Hosaka2023}
Y.~Hosaka, D.~Andelman, and S.~Komura,
``Pair dynamics of active force dipoles in an odd-viscous fluid,''
\textit{Eur. Phys. J. E} \textbf{46}, 18 (2023).

\bibitem{HosakaKomuraAndelman2021}
Y.~Hosaka, S.~Komura, and D.~Andelman,
``Nonreciprocal response of a two-dimensional fluid with odd viscosity,''
Phys.\ Rev.\ E \textbf{103}, 042610 (2021).
\bibitem{hosaka2026}
A.\ Daddi-Moussa-Ider, Y.\ Hosaka and S.\ Komura,
\emph{Exact response functions for a compressible thin fluid layer with odd viscosity},
arXiv:2602.18136 (2026).

\bibitem{KrishnanSamanta2026}
S.~Krishnan and R.~Samanta,
\emph{Hamiltonian active particles in incompressible fluid membranes},
arXiv:2512.03609 [cond-mat.soft] (2026).




\bibitem{Shoham2023}
Y.~Shoham and N.~Oppenheimer,
\emph{Hamiltonian dynamics and structural states of two-dimensional active particles},
Phys.\ Rev.\ Lett.\ {\bf 131}, 178301 (2023).



\end{thebibliography}
\end{document}